\documentclass[apj]{emulateapj}
\usepackage{graphicx}
\usepackage{epstopdf}
\usepackage{natbib}
\usepackage{xspace}
\usepackage{color}

\newcommand{\ggd}     {IRAS~18162--2048\xspace}
\newcommand{\kms}     {~km~s$^{-1}$\xspace}
\newcommand{\msun}    {~M$_{\sun}$\xspace}
\newcommand{\lsun}    {~L$_{\sun}$\xspace}
\newcommand{\formal}  {H$_2$CO~3$_{1,2}$--2$_{1,1}$\xspace}
\newcommand{\co}      {C$^{17}$O~2--1\xspace}
\newcommand{\lsbi}    {SO~$7_8$--$7_7$\xspace}
\newcommand{\lsbii}   {SO$_2$~16$_{3,13}$--16$_{2,14}$\xspace}
\newcommand{\lsbiii}  {SO$_2$~17$_{6,12}$--18$_{5,13}$\xspace}
\newcommand{\lsbiv}   {SO~$5_5$--$4_4$\xspace}
\newcommand{\usbii}   {SO$_2$~13$_{2,12}$--13$_{1,13}$\xspace}
\newcommand{\usbiv}   {SO$_2$~20$_{2,18}$--19$_{3,17}$\xspace}
\newcommand{\usbv}    {SO$_2$~6$_{4,2}$--7$_{3,5}$\xspace}
\newcommand{\cmd}   {~cm$^{-2}$\xspace}
\newcommand{\cmt}   {~cm$^{-3}$\xspace}
\newcommand{\vlsr}   {~$v_{\rm LSR}$\xspace}

\begin{document}

\title{A rotating molecular disk toward \ggd,\\ the exciting source of
HH~80--81}

\author{M. Fern\'andez-L\'opez\altaffilmark{1,6}}
\author{J.M. Girart\altaffilmark{2}}
\author{S. Curiel\altaffilmark{1}}
\author{Y. G\'omez\altaffilmark{3}}
\author{P.T.P. Ho\altaffilmark{4,5}}
\author{N. Patel\altaffilmark{5}}

\altaffiltext{1}{Instituto de Astronom{\'\i}a, Universidad Nacional Aut\'onoma de M\'exico
(UNAM), Apartado Postal 70-264, 04510 M\'exico, DF, M\'exico;
manferna@gmail.com, scuriel@astroscu.unam.mx}
\altaffiltext{2}{Institut de Ciencies de l'Espai, (CSIC-IEEC),Campus UAB, Facultat de Ciencies, Torre C5-parell 2, 08193 Bellaterra, Catalunya, Spain; girart@ieec.cat}
\altaffiltext{3}{Centro de Radioastronom\'{\i}a y Astrof\'{\i}sica, UNAM, Apartado Postal 3-72, Morelia, Michoac\'an 58089, M\'exico; y.gomez@astrosmo.unam.mx}
\altaffiltext{4}{Academia Sinica Institute of Astronomy and Astrophysics, P.O. Box 23-141, Taipei 10617, Taiwan}
\altaffiltext{5}{Harvard-Smithsonian Center for Astrophysics, 60 Garden Street, Cambridge, MA 02138, USA}
\altaffiltext{6}{Current address: Department of Astronomy, University of Illinois at Urbana--Champaign, 1002 West Green Street, Urbana, IL 61801, USA}

\section*{ABSTRACT}
We present several molecular line emission arcsec and subarcsec observations
obtained with the Submillimeter Array (SMA)  in the direction of the massive
protostar \ggd, the exciting source of HH~80--81. 

The data clearly indicates the presence of a  compact
(radius$\approx425$--850~AU) SO$_2$ structure, enveloping the more compact
(radius$\lesssim$150~AU)  1.4~millimeter dust emission  (reported in a previous
paper). The emission spatially coincides with the position of the prominent
thermal radio jet which  terminates at the HH~80--81 and HH~80N Herbig--Haro
objects.  Furthermore, the molecular emission is elongated in the direction
perpendicular to the axis of the thermal radio jet, suggesting a disk--like
structure. We derive a total dynamic mass (disk--like structure and protostar)
of 11--15\msun.  The SO$_2$ spectral line data also allow us to constrain  the
structure temperature between 120--160~K and the volume density
$\gtrsim2\times10^9$\cmt. We also find that such a rotating flattened system 
could be unstable due to gravitational disturbances. 

The data from C$^{17}$O line emission show a dense core within this
star--forming region. Additionally, the H$_2$CO and the SO emissions appear
clumpy and trace the disk--like structure, a possible interaction between a
molecular core and the outflows, and in part, the cavity walls excavated by the
thermal radio jet.

\section{INTRODUCTION}

\medskip

The powerful radiation from high-mass protostars was the main obstacle for 
understanding the formation of massive stars (see e.g. \citealt{1971Larson,
1974Kahn,1977Yorke, 1994Beech}; see also the recent review of
\citealt{2007Zinnecker}).  It was thought that the strong radiation pressure
from the protostar would stop the accretion, hence preventing the formation of
stars with masses over  $\sim$10~M$_{\sun}$. Although this theoretical  problem
can be overcome (e.g., \citealt{1987Wolfire,1989Nakano,  1996Jijina, 2002Yorke,
2000Norberg, 2003McKee, 2005Krumholz}), it  remains yet unclear how these stars
are actually formed. There are several hypotheses for the formation process.
Perhaps the most important ones are:  (1) large accretion rates (3 or 4 orders
of magnitude greater than those observed in low-mass protostars) through
circumstellar disks (\citealt{1995Walmsley, 1996Jijina, 2003McKee,
2005Zhang,2007Banerjee}),  (2) competitive accretion between small protostars
of the same cluster, that  could result in eventual mergers
(\citealt{1998Bonnell,2008Clarke,2011Moeckel,2011Baumgardt}), and   (3)
accretion of ionized material, even after the formation of a compact HII region
generated by the protostar (\citealt{2002Keto, 2006Keto}).

\begin{deluxetable*}{lcrlrrcc}
\tablewidth{0pc}
\tablecaption{Spectral observations}
\tablehead{
\colhead{Transition} & \colhead{$\nu_{line}$} & \colhead{$E_u$} &
\colhead{Data} & \multicolumn{2}{c}{Synthesized Beam} & \colhead{RMS}  &
\colhead{Figures}\\
\colhead{} & \colhead{(GHz)} & \colhead{(K)} & \colhead{} &
\colhead{$\arcsec\times\arcsec$} & \colhead{$\degr$} &
\colhead{(Jy~beam$^{-1}$)} & \colhead{} \\
}
\startdata
\lsbii   & 214.689380 & 147.9 &  LR & $8\farcs3\times3\farcs1$ & 34.2 & 0.065 
& \ref{Fspec_so2} \\
         &            &       & ROB & $1\farcs3\times0\farcs9$ & 59 & 0.085 &
\ref{Fcubos} \\
         &            &       &  HR & $0\farcs7\times0\farcs4$ & 13.2 & 0.085
& \ref{Fmoms} \\
\lsbiii  & 214.728330 & 229.1 &  LR & $8\farcs3\times3\farcs1$ & 34.2 & 0.065 &
\ref{Fspec_so2} \\
\usbv    & 223.883569 &  58.6 &  LR & $7\farcs9\times3\farcs0$ & 34.2 & 0.065 &
\ref{Fspec_so2} \\
\usbiv   & 224.264811 & 207.9 &  LR & $7\farcs9\times3\farcs0$ & 34.2 & 0.065 &
\ref{Fspec_so2} \\
\usbii   & 225.153702 &  93.1 &  LR & $7\farcs9\times2\farcs9$ & 34.2 & 0.065 &
\ref{Fspec_so2} \\
         &            &       & ROB & $1\farcs2\times0\farcs9$ & 58 & 0.070 &
\ref{Fcubos} \\
         &            &       &  HR & $0\farcs7\times0\farcs4$ & 12.9 & 0.080 &
\ref{Fmoms} \\
\lsbi    & 214.357004 &  81.3 &  LR & $8\farcs3\times3\farcs1$ & 34.2 & 0.065 &
\ref{Fspec_so2} \\
\lsbiv   & 215.220653 &  44.1 &  LR & $8\farcs2\times3\farcs1$
& 34.2 & 0.070 & \ref{Fspec_so2} \\
         &            &       & ROB & $1\farcs2\times0\farcs9$ & 58 & 0.080 &
\ref{Fcubos} \\
         &            &       & TAP & $4\farcs0\times2\farcs2$ & 36.7 & 0.080 
& \ref{Fsocube},\ref{Fmomso},\ref{Fespec} \\
\co      & 224.714385 &  16.2 &  LR & $7\farcs9\times2\farcs9$ & 34.2 & 0.075
& \tablenotemark{(a)} \\
         &            &       & TAP & $4\farcs0\times2\farcs1$ & 37.8 & 0.080 
& \ref{Fmomusb3} \\
\formal  & 225.697775 &  33.5 &  LR & $7\farcs9\times2\farcs9$ & 34.2 & 0.090 &
\ref{Fspec_so2} \\
         &            &       & ROB & $1\farcs2\times0\farcs9$ & 58 & 0.100 &
\ref{Fcubos} \\
         &            &       & TAP & $4\farcs0\times2\farcs1$ & 37.2 & 0.095 
& \ref{Fformalcube},\ref{Fmomformal},\ref{Fespec} \\
\enddata 
\tablecomments{Data extracted from JPL catalogue. The fourth column
\textit{Data} is a code giving the resolution of each map. LR means low angular
resolution data, HR, high angular resolution data, ROB, combined images with the
robust set to 0.3 and TAP, combined images with a taper applied and restricted
in the uvplane (see text, section \S2).}
\tablenotetext{(a)}{The compact configuration data of the \co is not shown
in any figure. However, it is used to obtain the velocity, density and
mass of the dense core of \ggd (see section \S 3.1).}
\label{Tsummary}
\end{deluxetable*}

\begin{figure}
\epsscale{1}
\plotone{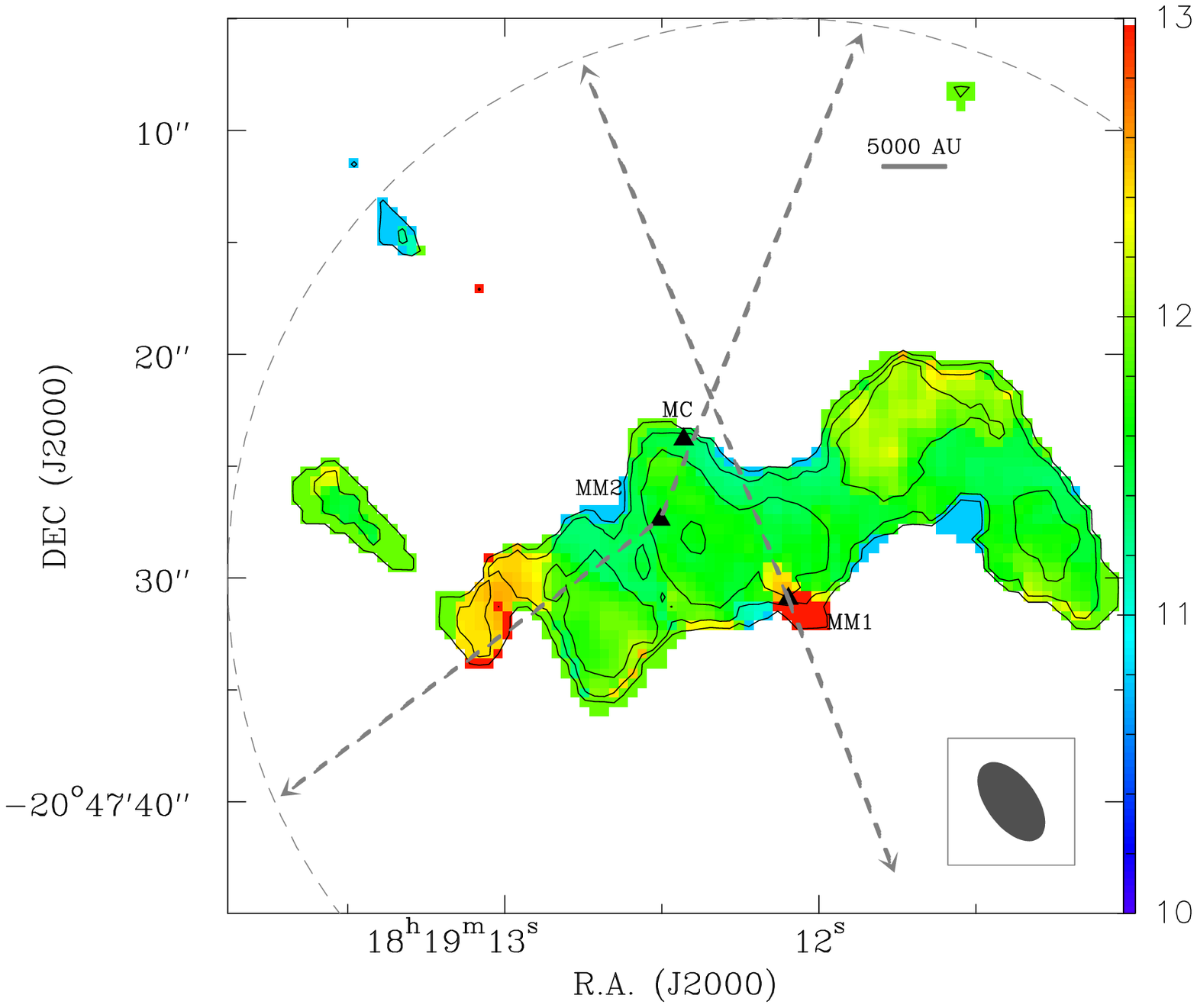}
\caption{Contour image of the zero order moment superposed on the first order 
moment colour image of the \co line emission (combined compact and extended 
configuration data with a Gaussian uv taper) toward \ggd. Contour levels 
are -9, -4, 4, 9, 15, 30, 45, 60, 75 and 90
times  0.10~Jy~beam$^{-1}$\kms, the rms of the image. The units of the scale bar
are in \kms. The molecular emission is integrated over velocities between 9 and
18\kms. The triangles mark the position of the millimeter sources MM1 and MM2,
and the position of the molecular core, MC. The dashed arrows show the
directions of the NW and SE outflows which seem to originate at the MM2 position
and also the NE-SW thermal radio jet, launched from MM1. The synthesized
beamsize is shown at the bottom right corner and the primary beam is indicated
by the dashed circle.
}
\label{Fmomusb3}
\end{figure}

\begin{figure}
\epsscale{1}
\plotone{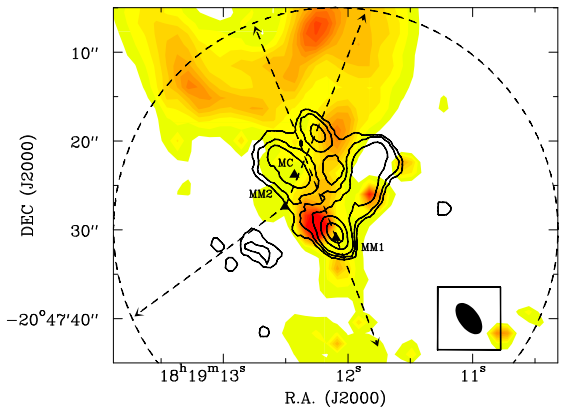}
\caption{Contour image of the zero order moment of the \lsbiv line emission 
(combined compact and extended configuration data with a Gaussian uv taper) 
superimposed on a K--band near infrared image of the reflection 
nebula in the \ggd region 
(courtesy of Thomas Geballe; see \citealt*{1992Aspin}). The infrared 
polarimetric study carried out by \cite{1987Yamashita} 
demonstrated that the scattered light from the infrared reflection nebula 
occurs on grains at the walls of a parabolic cavity rather than on
grains inside the whole outflow lobe.
Contour levels are -9, -4, 4, 9, 15, 30, 45, 60, 75 and 90 times 
0.17~Jy~beam$^{-1}$\kms, the rms of the image. Symbols are as in Fig.
\ref{Fmomusb3}.
}
\label{Fmomso}
\end{figure}

\begin{figure*}
\epsscale{1}
\plotone{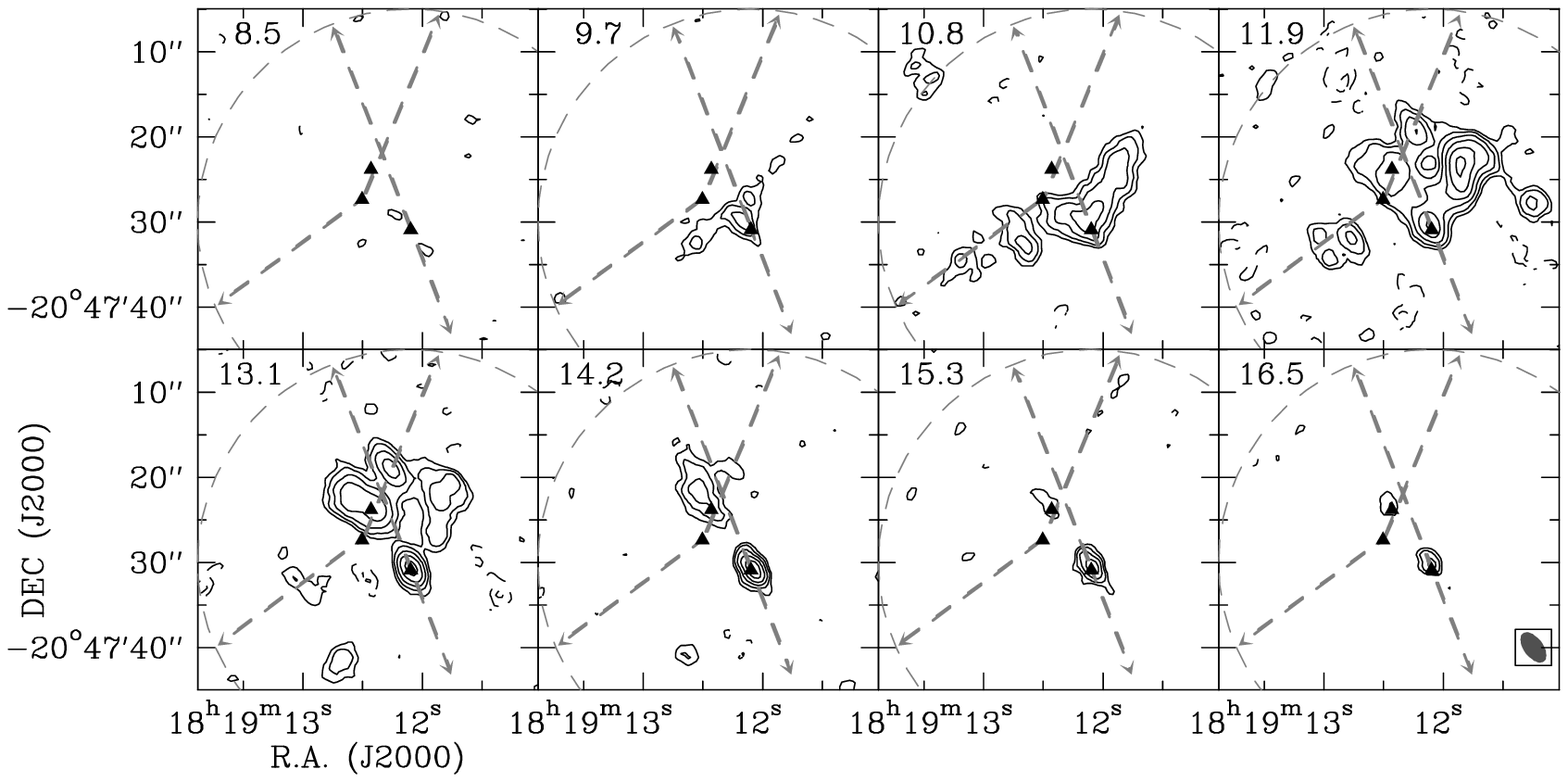}
\caption{Channel velocity image of the \lsbiv extended emission (combined 
compact and extended configuration data with a Gaussian uv taper)
from the central region of \ggd. The emission is averaged in
velocity bins of 1.14\kms. The velocity of each channel is indicated in the top
left corner of each panel in \kms units. The cloud velocity is at about
11.8\kms. The contour levels are -6, -4, 4, 6, 9, 13, 18, 24, 31, 39, 48 and 58
times 0.08~mJy~beam$^{-1}$, the rms of the image. Symbols are as in Fig.
\ref{Fmomusb3}.
The synthesized beam size is shown in the bottom left-hand corner of the most
redshifted channel. The blueshifted emission appears to trace mainly the walls
of the cavity excavated by the radio jet from MM1. At the position of MM1 the
image shows emission probably coming from the rotating disk/ring, absorbed in
the bluest channels by the extended emission. The redshifted channels show also
compact emission toward the position of MC.
}
\label{Fsocube}
\end{figure*}

\begin{figure}
\epsscale{0.7}
\plotone{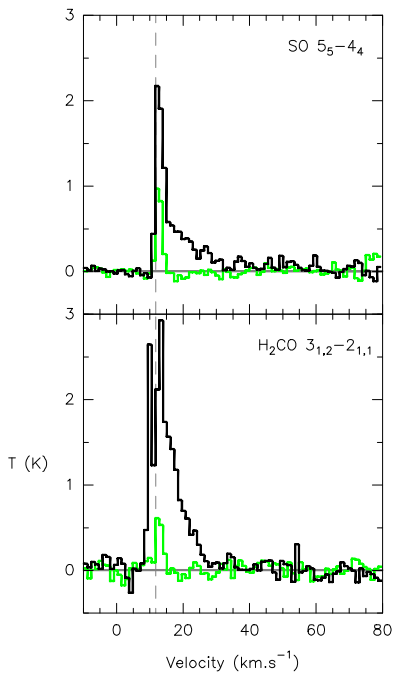}
\caption{\lsbiv (top panel) and \formal (bottom panel) spectra toward the center
position of MC (black line). Spectra toward the clump located to the north--west 
of MC (green line) are included to compare with that of MC. The dotted line marks 
the systemic 
velocity (11.8\kms). The intensity is in Kelvin. The spectra are extracted 
from the combined compact and extended configuration data with a Gaussian uv taper.
}
\label{Fespec}
\end{figure}

\begin{figure}
\epsscale{1}
\plotone{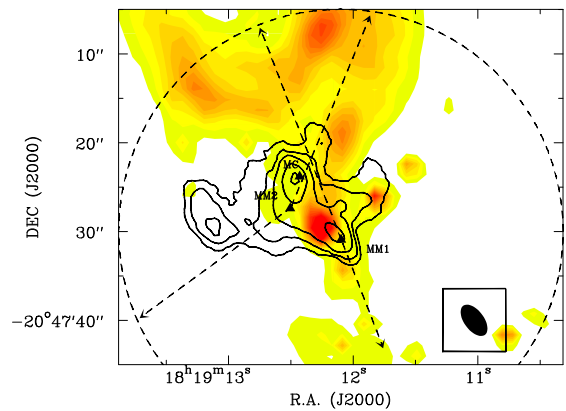}
\caption{Contour image of the zero order moment of the \formal line emission 
(combined compact and extended configuration data with a Gaussian uv taper) 
overlaid with a K--band near infrared image 
of the \ggd region (as shown in Fig. \ref{Fmomso}). 
Contour levels are -9, -4, 4, 9, 15, 30, 45, 60, 75 and 90 times 
0.33~Jy~beam$^{-1}$\kms, the rms of the image. Symbols are as in Fig.
\ref{Fmomusb3}.
}
\label{Fmomformal}
\end{figure}

\begin{figure*}
\epsscale{1}
\plotone{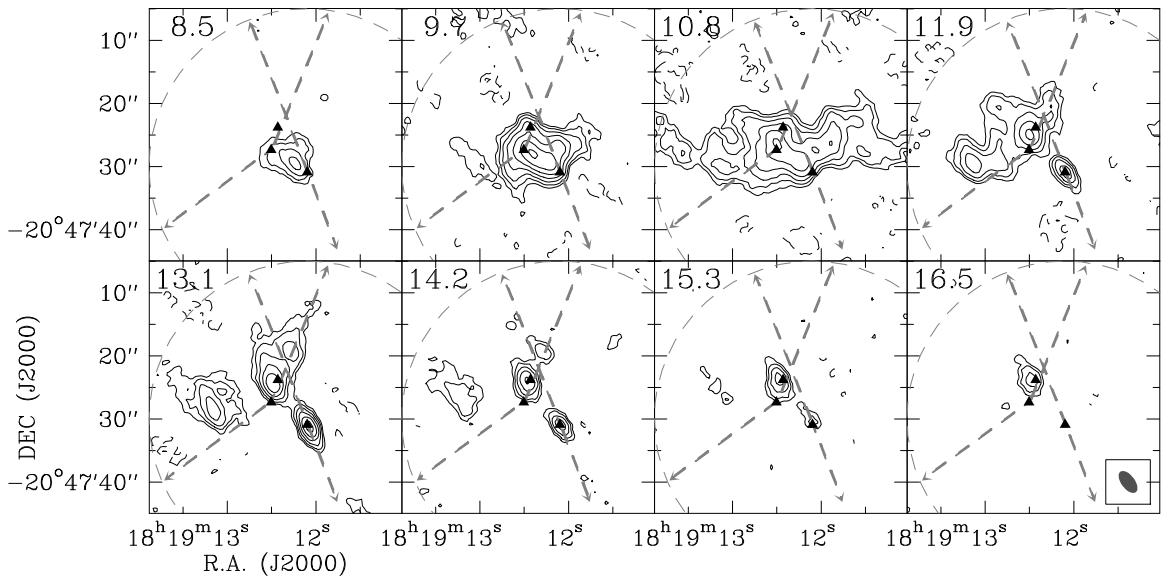}
\caption{Channel velocity image of the \formal extended emission (combined 
compact and extended configuration data with a Gaussian uv taper). The 
emission is averaged in velocity bins of 1.14\kms.
The velocity of each channel is indicated at the top left corner of each panel
in \kms units. The cloud velocity is at about 11.8\kms. The contour levels are
-6, -4, 4, 6, 9, 13, 18, 24, 31, 39, 48 and 58 times 0.095~mJy~beam$^{-1}$, the
rms of the image. Symbols are as in Fig. \ref{Fmomusb3}.
}
\label{Fformalcube}
\end{figure*}

In recent years, several works have provided evidence for the presence of 
collimated jets (HH~80-81, \citealt{1993Marti}; Cep~A, \citealt{2006Curiel}; 
IRAS~23139, \citealt{2006Trinidad}; IRAS~16547, \citealt{2008Rodriguez}), 
molecular outflows (see a summary in \citealt{2005Zhang,2007Arce}), and
flattened  structures of dust and gas (some with rotation and some with infall
signposts;  e.g., Cep~A, \citealt{2005Patel}; IRAS~20126,
\citealt{2005Cesaroni}; AFGL~2591, \citealt{2006VanderTak}; W51 North,
\citealt{2008Zapata};  IRAS~16547, \citealt{2009Franco-Hernandez}; W33A-MM1,
\citealt{2010GalvanMadrid}), surrounding high-mass protostars.  The evidence
leads to the interpretation that massive star formation is analogous  to
low-mass star formation, that is, via accretion from a flat rotating disk, with
a jet of ionized material, and with an associated molecular outflow. However,
the physical  characteristics of the possible accretion disks and their
associated outflows  have not been well characterized yet. Hence, the formation
process of massive stars  remains an open issue. There are several problems
that hamper the analysis of massive star formation regions (MSFRs).  First,
massive protostars form in clusters, which together with projection effects,
make these regions difficult to interpret. Moreover, the  MSFRs are found at
larger distances than low-mass star formation regions, with  typical distances
of 2-5~kpc. Thus, to get an insight of the closest regions to the protostars,
where accretion disks are expected, the highest angular resolution of the
current  telescopes is needed. Even so, most of the dust and gas structures
detected around high-mass protostars are $\sim1000$~AU. This is the smallest
diameter that an interferometer can resolve with a 1$\arcsec$ angular
resolution at those distances. At these scales, such structures could harbour
systems of several protostars, that could be forming high mass stars by other
mechanisms such as mergers. In fact, the nearest massive protostars (Orion
BN/KL and Cepheus A HW2), show complex scenarios, with a possible merger of
protostars (Orion BN/KL, \citealt{2009Zapata}), or close passages between
protostars (Cep~A HW2, \citealt{2009Cunningham}). Therefore, with the current
interferometers, only a few MSFRs may be studied with adequately high angular
resolution.

To demonstrate the existence of a disk around a massive protostar, it is not
enough to detect its (sub)millimeter dust continuum emission;  its kinematics
must also be measured via the molecular  line emission associated with the
gas--phase chemistry developed  after the evaporation of the molecules from the
ice mantles  (\citealt{1998Hatchell,2007Cesaroni}). However, searching for
molecular tracers of disks in MSFRs is a complex task. There are molecules that
show emission from the envelope and the disk simultaneously. Other molecules
show optically thick emission, thus complicating the kinematic study of the
disk.  On the other hand, S-bearing species (such as H$_2$S, SO, SO$_2$, CS,
OCS...) could be intimately linked with the evaporation process of the disk
surface (\citealt{1997Charnley,1998Hatchell,2003VanderTak,2005MartinPintado}),
becoming  good tracers of the dynamics of the innermost parts of the high mass
protostars.  Recently, several papers have been published on the detection of
S-bearing species in disks and other warm gas-structures of MSFRs 
(\citealt{2006VanderTak,2007JimenezSerra,2009Klaassen,2009Franco-Hernandez,
2009Zapata}). In particular, SO$_2$ transitions, ubiquitous within the
(sub)millimeter range, show a very compact nature, suggesting their close 
association with circumstellar structures.

The GGD27 complex is an active star forming region located at a distance of 
1.7~kpc. It shows a spectacular (5.3~pc long)  and highly collimated thermal
radio jet (\citealt{1993Marti,1995Marti,1998Marti}), embedded within a powerful
CO bipolar outflow (\citealt{1989Yamashita,2004Benedettini}). The jet has a
position angle of $\sim21\degr$ (\citealt{1993Marti,1999Marti}) and ends at two
southern and very bright Herbig-Haro objects (HH~80 and 81, originally
discovered by \citealt{1988Reipurth}) and at a radio source to the north (HH~80
North,  \citealt{1994Girart}). Linear polarization has been detected in the
radio emission  from this jet, indicating the presence of a magnetic field
coming from the disk (\citealt{2010Carrasco}).The central part of the radio jet
has a bright  far-infrared counterpart (\ggd), that implies the presence of a
luminous young star ($\sim2\times10^4$~L$_{\sun}$) or a cluster of stars
(\citealt{1992Aspin,1997Stecklum}).

Submillimeter and millimeter wavelength observations of the central part of the 
thermal radio jet have revealed two dusty sources, MM1 and MM2, separated by about
7$\arcsec$ (\citealt{2003Gomez, 2009Qiu}), which are apparently in very 
different evolutionary stages (\citealt{2011Fernandez}, hereafter 
Paper I). MM1, the south-western source, coincides with the origin of the 
thermal radio jet and has a dust temperature of 109~K. 
It has a weak extended envelope, possibly surrounding a very compact
(R$\lesssim$150~AU) disk-protostar system. 
The mass of the disk derived from the continuum emission is $\sim4.1$~\msun.
Such a massive disk could have an extremely high accretion rate
($10^{-3}-10^{-2}$\msun~yr$^{-1}$). However, the case for an accretion
disk in MM1 has not been unambiguously confirmed yet. The bolometric luminosity
obtained from fitting the SED of high angular resolution data is
$\geq3300$\lsun, similar to that of a B1 Zero Age Main Sequence (ZAMS) star.
However, the massive disk of MM1 (compared to those of low-mass
protostars), the presence of the powerful outflow and the youth of the protostar
(it has not developed a compact H~II region yet), indicate that MM1 could
become a B0 or a more massive star. MM1 is also associated with compact emission
from several high density tracers, such as CS (\citealt{1991Yamashita}), SO
(\citealt{2003Gomez}), CH$_3$OH, CH$_3$CN, H$_2^{13}$CO, OCS, HNCO and SO$_2$
(\citealt{2009Qiu}). 

The physical characteristics of MM2 are typical of the Class 0 low--mass
protostars (see e.g., \citealt{1993Andre}), except for its total  mass of at
least 11~M$_{\sun}$ (4-5 times more mass than a typical low-mass Class 0
source, see e.g. \citealt{2006Girart,2009Rao}). MM2 appears as an extended
source  when observed with low angular resolution at 1.4~mm, but it is
separated into two compact sources when observed with high angular resolution
(Paper I). The stronger component in MM2 spatially coincides with very weak
free-free emission, suggesting that it could be a high/intermediate mass
protostar. The weaker component in MM2 could be a pre-protostellar core.
Previous CO observations (\citealt{2009Qiu}) have associated the position of
MM2 with the origin of a young outflow running to the east. In addition,
another possible CO outflow, running north-west of MM2 could also have 
originated close to this position.

There is also a possible molecular core (hereafter MC, \citealt{2009Qiu})
located about $4\arcsec$ to the north-west of MM2. Although MC is traced by
several molecules (CH$_3$CN, H$_2^{13}$CO, OCS y HNCO, \citealt{2009Qiu}, and a
Class I CH$_3$OH maser, \citealt{2004Kurtz}), it is not associated with a
bright and compact millimeter source (Paper I). The diverse line emission from
MC has been interpreted by \cite*{2009Qiu} as the emission from a hot core
warmed by the radiation of a young and massive protostar. 

Here we present an analysis of the molecular emission of several lines (mostly
S-bearing species) detected in the central region of \ggd using the
Submillimeter Array\footnote{The Submillimeter Array is a joint project between
the Smithsonian Astrophysical Observatory and the Academia Sinica Institute of
Astronomy and Astrophysics and is funded by the Smithsonian Institution and the
Academia Sinica.} (SMA; \citealt{2004Ho}). The SMA allowed us to detect several
transitions simultaneously, some of which seem to trace a circumstellar
disk/ring around MM1, with subarcsecond resolution  ($\sim0\farcs5$),
permitting a beam size of $\sim850$~AU at the source distance.  This resolution
is necessary in order to perform a reliable kinematical analysis of the motions
of the circumstellar disks at distances greater than 1--2~kpc.

The continuum emission of the observations was presented in a recent
publication (Paper I). In section 2 we briefly describe the observations, while
in section 3 the results of the observations are given. Section 4 contains the
analysis of the column density and the  temperature of the disk/ring of MM1,
followed by a discussion in section 5. Finally, in section 6 we draw our main
conclusions.

\section{OBSERVATIONS}
The SMA observations were taken during two epochs: on 2005 August 24
in the compact configuration (giving a synthesized beam of $8\farcs1\times
3\farcs0$) and on 2007 May 27 in the very extended configuration (giving a
synthesized beam of $0\farcs7\times 0\farcs4$). During both epochs the receiver
was tuned at 215/225~GHz. The phase center of the telescope was RA(J2000.0)$=
18^h 19^m 12\fs1$ and DEC(J2000.0)$= -20\degr47\arcmin 30\farcs0$, and the 
correlator provided a spectral resolution of about 1.14~km~s$^{-1}$ at
the observed frequency (with 0.81~MHz of channel width). The continuum data were
edited and calibrated using Miriad (\citealt{1995Sault}). The flux uncertainty
was estimated to be $\sim20$\%. The detailed description of the observing setup
as well as the calibration process of the continuum emission data can be found
in Paper I. 

Nine emission lines were detected with the compact configuration (Table
\ref{Tsummary}), four of which were also detected with the very extended
configuration (\lsbii, \usbii, \lsbiv, \formal). The solutions of the
self--calibration performed on the continuum data were transferred to the line
data, and then the molecular lines were imaged, cleaned, restored, and analyzed
using the Miriad and AIPS (developed by NRAO) packages. The  line data were
also corrected for the half-channel error in the SMA velocity labeling
discovered in 2007 November. After that, we used a channel separation of
1.136\kms for all the lines. The calibration was performed in the same manner
for the compact and very extended configuration data, except for the flux
calibration step (see Paper I).  Finally, the two data sets (compact and very
extended configurations) were combined for all the lines giving  maps with an
intermediate angular resolution (hereafter combined data). The average rms
noise level is  $\sim$70~mJy~beam$^{-1}$ in the line channel maps obtained with
the compact configuration data, $\sim$85~mJy~beam$^{-1}$ in those obtained with
the very extended configuration, and $\sim$85~mJy~beam$^{-1}$ in those obtained
with the combined data, all with a natural weighting.  Throughout the paper,
all the velocities are given in LSR.

We use the high angular resolution data (very extended configuration) to
analyze the disk kinematics and the low angular resolution data (compact
configuration) to extract the spectra and estimate the disk temperature and
column density. In addition, we obtain two different sets of  images using the
combined data to examine the extended emission with different angular
resolutions. First, we set the robust parameter to 0.3. It yields a synthesized
beam of about $1\farcs2\times0\farcs9$ (Table \ref{Tsummary}). Second, we apply
a taper of 182~k$\lambda$ and restrict the visibilities up to 200~k$\lambda$,
thus improving the sensitivity to the extended emission. The synthesized beam
of the tapered data is about $4\farcs0\times2\farcs1$ (see Table
\ref{Tsummary}). 

\section{RESULTS}
\begin{deluxetable*}{lccccccc}
\tablecolumns{8}
\tablewidth{0pc}
\tablecaption{Gaussian fits to the low angular resolution data toward MM1.}
\tabletypesize{\small}
\tablehead{
\colhead{Line} & \colhead{$S_{\mathrm{peak}}$} & \colhead{V$_{\mathrm{LSR}}$} &
\colhead{FWHM\tablenotemark{(a)}} & \colhead{Gaussian area}\tablenotemark{(b)} &
\colhead{RMS\tablenotemark{(c)}} & \colhead{Zero-to-zero}   \\
\colhead{} &\colhead{K} & \colhead{km~s$^{-1}$} &  \colhead{km~s$^{-1}$} &
\colhead{K~km s$^{-1}$} & \colhead{K} & \colhead{km~s$^ {-1}$} \\
}
\startdata
\lsbii   & $1.32\pm0.03$ & $13.04\pm0.05$ & $5.9\pm0.2$ & $8.2\pm0.4$ & 0.03 &
5.5-19.0 \\
\usbii   & $1.33\pm0.05$ & $12.9\pm0.1$ & $6.0\pm0.4$ & $8.5\pm0.4$ & 0.07 &
6.5-18.0 \\
\lsbiii  & $0.56\pm0.05$ & $13.2\pm0.3$ & $5.7\pm0.4$ & $3.4\pm0.8$ & 0.06 &
8.5-20.0 \\
\usbiv   & $0.88\pm0.05$ & $12.8\pm0.2$ & $6.6\pm0.4$ & $6.2\pm0.7$ & 0.06 &
6.0-19.0 \\
\usbv   & $0.49\pm0.05$ & $13.4\pm0.2$ & $6\pm0.5$ & $3.3\pm0.7$ & 0.05 &
6.5-20.0 \\
%\tableline
\lsbiv                      & $2.29\pm0.04$ & $12.86\pm0.04$ & $5.65\pm0.05$ &
$13.8\pm0.5$ & 0.04 & 6.0-19.0 \\
\lsbi\tablenotemark{(d)}    & $0.5\pm0.1$ &  $9.8\pm0.3$ & $3\pm0.5$ &
$1.3\pm0.4$  & 0.07 & 6.5-18.0 \\
			    & $0.63\pm0.05$ & $13.8\pm0.3$ & $4\pm0.5$ &
$2.5\pm0.4$  &      &  \\
\formal\tablenotemark{(d)}  &  $3.6\pm0.1$ &  $9.72\pm0.05$ & $2.1\pm0.1$ &
$8.3\pm0.6$  & 0.10 & 5.0-17.0 \\
			    &  $3.7\pm0.1$ & $13.10\pm0.05$ & $2.6\pm0.2$ &
$6.56\pm0.6$  &      &  \\
\enddata 
\tablecomments{The spectra were obtained over a box (-8,-8,7,7) arcsecs. Gaussian
fits were carried out with a program based on the algorithm by
\citealt{2009Canto}.}
\tablenotetext{(a)}{Calculated from $\sigma_{v}\cdot 2\sqrt{\ln{2}}$.}
\tablenotetext{(b)}{It is obtained as $\sqrt{\pi}\cdot
S_{\mathrm{peak}}\cdot\sigma_{v}$.}
\tablenotetext{(c)}{Estimated from the free-line part of each spectrum.}
\tablenotetext{(d)}{A simultaneous two-gaussian fit was performed for this
transition.}
\label{Tlowspec}
\end{deluxetable*}

\begin{figure*}
\epsscale{1}
\plotone{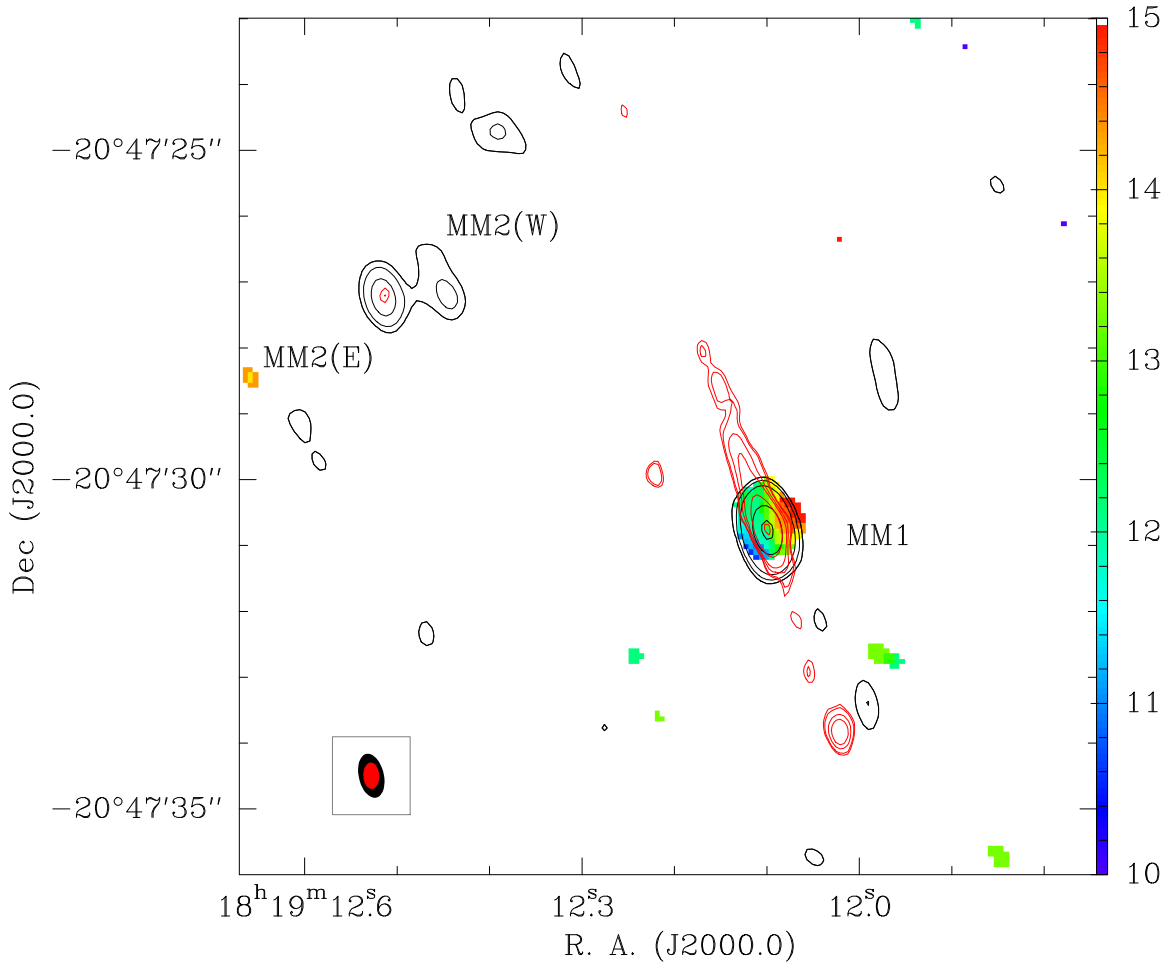}
\caption{Image of the VLA radio continuum emission at 3.5~cm
(red contours), SMA 1.4~mm continuum emission (black contours) and 
the SMA integrated emission of the \usbii line (colour scale). The contour
levels of the 3.5~cm map are 5, 6, 10, 15, 40 and 200 times 
8.8~$\mu$mJy~beam$^{-1}$, the rms of this image. 
The contour levels of the 1.4~mm image are -5, 3, 3, 5, 
10, 60 and 120 times 0.003~mJy~beam$^{-1}$, the rms of this image. 
The \usbii molecular emission is integrated over the velocity range 
7--19\kms, and the wedge panel in the right shows the colour scale 
intervals. In the bottom left corner the synthesized beams of the 
cm image (red ellipse) and the mm images (black ellipse) are shown.
}
\label{Fposter}
\end{figure*}

\subsection{C$^{17}$O emission}

Fig. 1 shows the integrated emission of the \co molecular line superposed on
the first order moment. The C$^{17}$O emission extends mainly along a
south--east  to north--west lane of $\sim23\arcsec\times10\arcsec$ 
($\sim0.2$~pc~$\times$~0.1~pc) and a position angle of $\sim120\degr$,
approximately perpendicular to the thermal radio jet (position angle of
21$\degr$).  The two main peaks of the \co emission appear on both sides of the
thermal radio jet (see Fig. \ref{Fmomusb3}), with the strongest peak at a
position between the two dominant (sub)millimeter sources, MM1 and MM2. MM1
(the source from which the radio jet is launched), is located at the bottom
center of the \co lane emission. 

Fig. \ref{Fmomusb3} also shows that the v$_{\rm LSR}$ of the line is almost
constant along the lane structure.  The line is narrow and has a symmetric
profile, with a full width at half maximum (FWHM) of 2.8\kms$\pm0.03$. A
Gaussian fit to this line profile gives  also a central velocity of
11.8$\pm$0.03\kms. In what follows we will consider this value as the systemic
velocity. This velocity agrees well, to  within our spectral resolution, with
that measured by \cite{2003Gomez}  (12.2$\pm$0.1\kms) from the emission of the
ammonia core.

The critical density of the C$^{17}$O 2-1 line is relatively low,  $n_{\rm
crit} \sim 10^4$cm$^{-3}$,so we expect that the emission detected by  the SMA
is likely thermalized. Assuming LTE conditions will yield a lower  limit for
the total mass of the extended  molecular cloud. We assume a C$^{17}$O
abundance of $4.7\times10^{-8}$ (\citealt{1982Frerking}) and an excitation
temperature of 24~K, equal to the T$_{\rm rot}$ found by \cite{2003Gomez} using
the NH$_3$ VLA observations. From equations B1 and B2 of \cite{2010Frau} and
for optically thin emission, we obtain an average molecular column density of
$7\times10^{22}$\cmd. Hence, we estimate a lower limit of 38\msun for the total
mass.

\subsection{SO emission}

The SMA correlator setting includes two SO transitions: \lsbi and \lsbiv. We
detect \lsbi only in the compact configuration due to a problem with the
correlator at the frequency of this line in  the very extended configuration. 
The \lsbi line emission from the compact configuration data appears as an
unresolved clump toward MM1. On the other hand, the emission from the  \lsbiv
line at the systemic velocity of the dense core shows a  clumpy distribution
mainly to the north of MM1, having a conical shape  which surrounds the north
lobe of the thermal radio jet (Figs. \ref{Fmomso} and  \ref{Fsocube}). Some of
the weaker clumps seem to coincide  with the direction of the north--west and
south--east outflows observed by \cite*{2009Qiu}.  The redshifted emission
(with respect to the cloud velocity) is compact and is mainly distributed
around MM1 and MC (Fig. \ref{Fsocube}). The \lsbiv line profile at the position
of MC (top panel of Fig.  \ref{Fespec}) has associated a clear redshifted wing 
with emission spreading up to a few tens of km~s$^{-1}$.

\subsection{H$_2$CO emission}

The emission of the \formal line is more extended than that of the \lsbiv line,
resembling the distribution of the elongated \co emission  around the systemic
velocity (Fig. \ref{Fmomformal} and \ref{Fformalcube}).  Furthermore, at the
11.9 and 13.1\kms\ velocity channels the emission  appears to follow the
direction of the two outflows mentioned above. At redshifted velocities (\vlsr\
$\ga 13.1$\kms) the emission is mainly distributed around the north lobe of the
radio jet, similarly to that observed in the SO redshifted emission.  Contrary
to the SO, the \formal presents emission around  MM 2. On the other hand and 
as occurs with the SO, the \formal emission is enhanced at the positions of MM1
and MC.  Toward MC, the \formal line shows a high velocity red wing, more
prominent   than that observed in \lsbiv line profile (bottom panel of Fig.
\ref{Fespec}).

\subsection{SO$_2$ emission}

Five SO$_2$ lines were detected with the SMA in its compact configuration 
(Table \ref{Tlowspec}). All of them show unresolved emission at the origin of
the thermal radio jet, which coincides with the MM1 (sub)millimeter continuum
source  (Fig. \ref{Fposter}). \lsbii and \usbii were also detected by the SMA
in its  very extended configuration. The other  SO$_2$ lines were undetected
due to their low intensity, their (probably) partially resolved nature and the
lower sensitivity of the high angular resolution data. The panels a) and b) of
Fig. \ref{Fcubos} show the velocity channel cubes of the \lsbii and \usbii
lines  (images of the combined data with robust 0.3). The molecular structure
at the position of MM1 appears at the  8.7\kms to 15.6\kms velocity channels,
while the peak velocity is  \vlsr$\simeq 13.0$\kms (see Table~\ref{Tlowspec}).
The peak velocity of the molecular structure is therefore redshifted with
respect to  the large scale dense core velocity by $\sim1.2$\kms. The
blueshifted channels (8.7 to 11.0\kms) are seen to the east, while the
redshifted channels (14.4 and 15.6\kms) are seen to the west with respect to
the thermal radio jet and the 1.4~mm continuum peak position. 

\begin{deluxetable*}{llllclll}
\tablecolumns{6}
\tablewidth{0pc}
\tablecaption{Size and kinematics of the SO$_2$ disk. From the high angular
resolution data.}
\tabletypesize{\small}
\tablehead{
\colhead{Line} & \multicolumn{3}{c}{Deconvolved Size} &
\colhead{R\tablenotemark{(a)}} & \colhead{V$_{\rm disp}$\tablenotemark{(b)}} &
\colhead{$\Theta_{\rm disp}$\tablenotemark{(c)}} & \colhead{M$_{\rm
dyn}$\tablenotemark{(d)}} \\
\colhead{} & \colhead{$\arcsec$} & \colhead{$\arcsec$} & \colhead{$\degr$} &
\colhead{AU} & \colhead{\kms} & \colhead{\msun} & \colhead{\msun} \\
}
\startdata
\lsbii   & $0.88\pm0.08$ & $0.81\pm0.05$ & $130\pm13$ & $720\pm50$ & $8.6\pm0.6$
& $0.87\pm0.15$ & $15\pm5$ \\
\usbii   & $1.20\pm0.06$ & $0.81\pm0.05$ &  $165\pm6$ & $840\pm50$ & $6.7\pm0.6$
& $1.03\pm0.15$ & $11\pm3$ \\
\enddata 

\tablenotetext{(a)}{Equivalent radius of the disk. For instance
$R=\sqrt{b_{maj}\cdot b_{min}}/2$.}
\tablenotetext{(b)}{Maximum velocity dispersion from the 50\% contour of the p-v
diagrams. The uncertainty is half the velocity channel.}
\tablenotetext{(c)}{Maximum spatial offset from the 50\% contour of the p-v
diagrams. The uncertainty is half the minor synthesized beam axis, which is
roughly in the direction of the disk.}
\tablenotetext{(d)}{Obtained using the expression (see e.g.,
\citealt{2005Patel,2008Beuther,2008Zapata}):
$(M/M_{\sun})/\sin^2(i)=1.121\times10^{-3}\times((\Delta v /
km~s^{-1}) /2)^2\times(R/AU)$, where $i$ is the inclination of the disk with
respect to the plane of the sky, $\Delta v$ is the velocity difference measured
at the 50\% contour level and R is the radius fitted with the AIPS task IMFIT.}
\label{Tmdyn}
\end{deluxetable*}

The left panels of Fig. \ref{Fmoms} show the zero order moment (i.e.,
integrated  emission) and first order moment (i.e., integrated velocity
weighted by the intensity)  images built up from the \lsbii and \usbii high
angular resolution cubes.  The  molecular structure toward MM1 is compact and
its peak coincides with the position  of the thermal radio jet within
$0\farcs2$. The deconvolved sizes of the zero order  moment images of \lsbii
and \usbii are presented in Table \ref{Tmdyn}. Assuming a  distance of 1.7~kpc
to \ggd, the equivalent radius of the emission ($\sqrt{Area/\pi}$)  is less
than about 850~AU for both lines. The first order moment images clearly show 
the southeast--northwest velocity gradient nearly perpendicular to the radio 
jet axis. In fact, the position angle between the most redshifted and
blueshifted channels are about $\sim110\degr$ and $\sim130\degr$ for the \lsbii
and \usbii lines, respectively. In section \S 5.1 we argue that the SO$_2$
emission arises from  a disk/ring rotating structure.

\section{ANALYSIS}
\subsection{RADEX}\label{Sradex}

\begin{figure*}
\epsscale{1}
\plotone{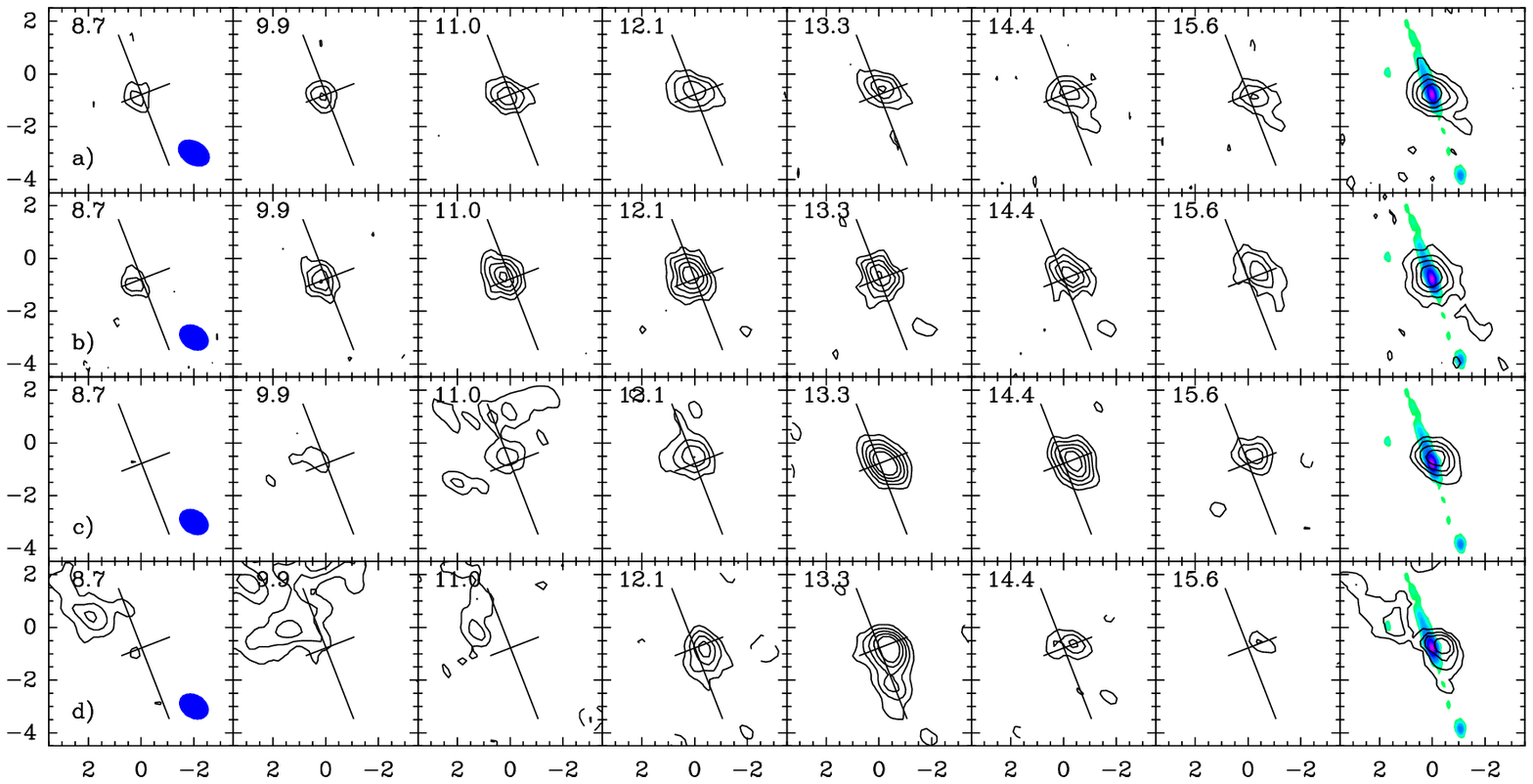}
\caption{SMA channel velocity images (combined data with robust 0.3) of several
lines toward the position of MM1, the source from which originates the
thermal radio jet. The lines are presented in the
following order: row a) \lsbii, row b) \usbii, row c) \lsbiv and row
d) \formal. The velocity channels are 1.14\kms wide and the central velocity
is indicated in the upper left corner of the channels. The disk velocity center
is 13.0\kms while the dense core velocity is 11.8\kms. Contours are -5, -3, 3, 5,
7, 9, 11 and 13 times the rms noise level of each image, given in Table
\ref{Tsummary}. The direction of the radio jet and the direction perpendicular
to it are marked by two straight lines in each channel. The last channel of each
row shows the zero order moment of each line (contours) superimposed on the 3.5~cm 
continuum radio jet image (grey scale). The
synthesized beam sizes appear as a blue ellipse in the first channel of each
row. It is evident that the peak of the emission moves from east to west on the
SO$_2$ velocity cubes.
}
\label{Fcubos}
\end{figure*}

\begin{figure}
\epsscale{1}
\plotone{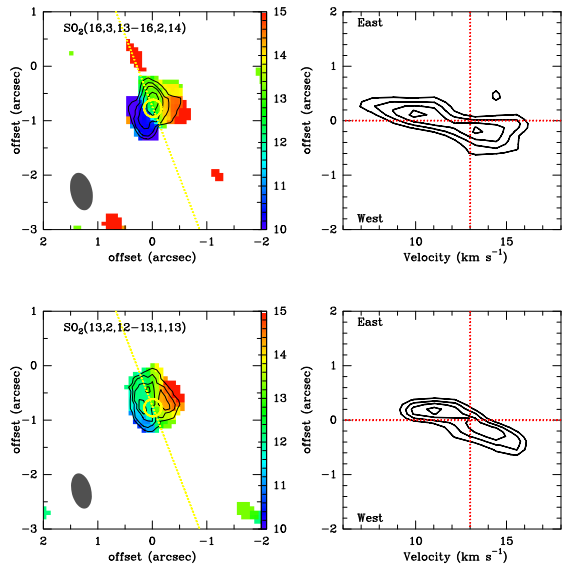}
\caption{\lsbii and \usbii zero (contours) and first order (colour scales)
moment images (left panels) toward MM1. Their corresponding
position--velocity diagrams with a perpendicular crosscut through the jet axis
(P.A.$=111\degr$) at the position of MM1 are shown in the right--hand panels. 
The contour levels for the zeroth order moments are -30, 30, 50,
70 and 90\% the flux peak, and the colour bar is in \kms. The emission is
integrated over the velocity range 7--19\kms for both lines. The yellow dotted
line marks the orientation of the radio jet. The yellow circle is centered at
the position of the millimeter peak continuum emission of MM1 and has the
equivalent radius of the upper limit of the size obtained from the 1.4~mm
millimeter image (Paper I). In the position--velocity diagrams (panels on the
right) the vertical dotted line marks the disk velocity. East is at the top and
west is at the bottom of both position--velocity diagrams. The contours are 50\%
to 95\% with a step of 15\% the peak flux.
}
\label{Fmoms}
\end{figure}

\begin{figure}[h]
\epsscale{1}
\plotone{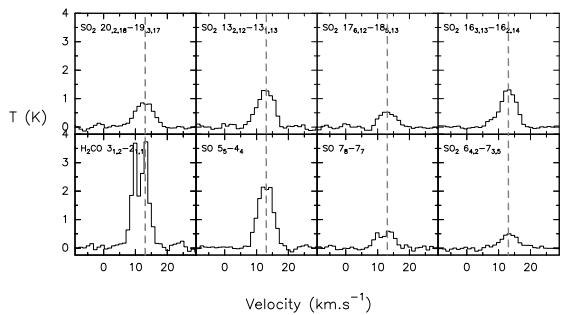}
\caption{Spectra of several transitions toward MM1 from low angular resolution
data. The dashed line is placed at the velocity of the disk peak emission
(13.0\kms on average). Intensity is in Kelvin.
}
\label{Fspec_so2}
\end{figure}

\begin{figure}
\epsscale{0.85}
\plotone{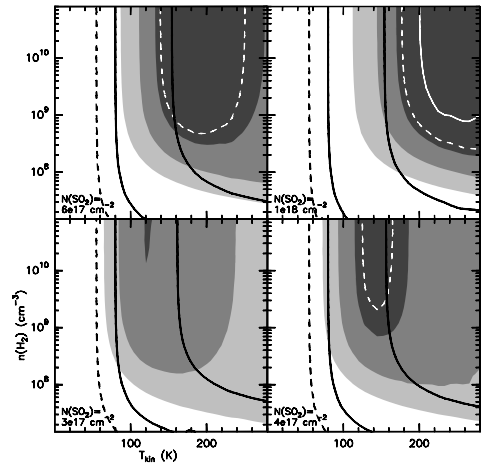}
\caption{Plot of the set of RADEX solutions in the $n$(H$_2$)-$T_{\rm kin}$
plane for the emission associated with the MM~1 disk for four different values
of the SO$_2$ column density (from bottom-left to top-right panels): 3, 4, 6 and
$10 \times 10^{17}$\cmd.
The grey image gives the values of the $\chi^2$ solutions for the SO$_2$ line
ratios (see \S~\ref{Sradex}). The increase of the grey tonalities (from lighter 
to darker) indicated a decrease in the $\chi^2$ values. The solid and dashed 
white contours show the 68\% and 99\% confidence region of the $\chi^2$.
The dashed black line shows the range of solutions for a brightness temperature
of the 13$_{2,12}$-13$_{1,13}$ line of 50~K, which is the value measured at the
high angular resolution map. The area between the two black solid line shows the
range of valid solutions for the 13$_{2,12}$-13$_{1,13}$ brightness temperature
if the SO$_2$ has an emitting size between $0\farcs5$ and $1\farcs0$ (see
\S~\ref{Sradex}). 
} 
\label{Radex}
\end{figure}

\begin{deluxetable}{ccc}
\tablecolumns{6}
\tablewidth{0pc}
\tablecaption{Size of the SO$_2$ disk. 
II. From high to low angular resolution ratio}
\tabletypesize{\small}
\tablehead{
\colhead{} & 
\colhead{SO$_2$} & 
\colhead{SO$_2$} 
\\
\colhead{Parameter} & 
\colhead{13$_{2,12}$-13$_{1,13}$} & 
\colhead{16$_{3,13}$-16$_{2,14}$} 
}
\startdata
$T_{\rm mb}^{\rm high}$ (K)	& $50\pm8$ 		& $53\pm10$ \\
$T_{\rm mb}^{\rm low}$ 	(K)	& $1.33\pm0.05$ 	& $1.32\pm0.03$ \\
$T_{\rm mb}^{\rm high}/T_{\rm mb}^{\rm low}$	& $37.6\pm6.2$ 	& $40.2\pm7.6$
\\
$\Omega$ (arcsec) 			& $0\farcs62\pm0\farcs10$ &
$0\farcs59\pm0\farcs11$ \\
$\Omega$ (AU) 				& $1054\pm170$& $1003\pm187$ \\
\enddata 
\label{Tradex}
\end{deluxetable}

In this section we estimate the physical characteristics of the MM1 disk, such
as volume density and temperature by means of RADEX modeling. We used RADEX to
simulate the line intensities of the five observed SO$_2$ transitions (Fig.
\ref{Fspec_so2}).  The RADEX code is a non-LTE molecular radiative transfer
code which assumes an isothermal homogeneous medium (\citealt{2007VanderTak}). 
This assumption is reasonable as a first approach to constrain the physical
properties of the gas traced by the SO$_2$ since the observed lines are
optically thick (see below).  We explored a range of values between 50 and
300~K in the kinetic temperature, between $10^7$ and $10^{11}$\cmt\ in the
volume density, $n$(H$_2$), and between $10^{14}$ and $10^{19}$\cmd\ in the
SO$_2$ column density, $N$(SO$_2$).  In order to constrain the physical
properties of the SO$_2$ emitting region  the next scheme was followed:

\begin{itemize}
\item 
We first computed the $\chi^2$ parameter (which is a measure of how well the
model fits the observations) for the line ratio of the 
16$_{3,13}$-16$_{2,14}$, 17$_{6,12}$-18$_{5,13}$, 20$_{2,18}$-19$_{3,17}$ and
6$_{4,2}$-7$_{3,5}$ SO$_2$ transitions with respect to the
13$_{2,12}$-13$_{1,13}$ SO$_2$ transition. This approach assumes that all the 
transitions trace the same region (the images from the combined data of these 
lines give similar deconvolved sizes all with deconvolved major axis 
$\lesssim3\arcsec$). Figure~\ref{Radex} shows in grey scale a set of the best 
solutions in the $n$(H$_2$)-$T_{\rm kin}$ plane for four different  values 
of the SO$_2$ column density. The best set of solutions (darkest areas of 
Fig.~\ref{Radex}) indicates that the SO$_2$ arises from hot and very dense 
molecular gas, as expected if it is associated with the MM1 disk. Indeed, the 
line ratios do not constrain well the physical values of the gas but rather 
yield lower limits:
$n$(H$_2$)$\ga 3\times10^8$\cmt\ and $T_{\rm kin} \ga 120$~K. In addition,
RADEX predicts a very large molecular column density of SO$_2$, $N$(SO$_2$)$\ga
3\times10^{17}$\cmd, indicating that the observed lines are optically thick. 
\item
As an additional constraint, we take into account the possible values of the 
SO$_2$ 13$_{2,12}$-13$_{1,13}$ and 16$_{3,13}$-16$_{2,14}$ brightness 
temperatures. As a lower limit we can use the main beam brightness temperature 
measured in the highest angular resolution maps, $T_{\rm mb} = 50$~K (it  is
shown in Figure~\ref{Radex} as a dashed line).  However, it is not enough to
further constraint the physical conditions of the gas associated with the
SO$_2$ emission. As a further constrain, the true brightness temperature of the
SO$_2$ can be estimated if the source size is known. The high angular
resolution maps of the 13$_{2,12}$-13$_{1,13}$  and 16$_{3,13}$-16$_{2,14}$
integrated emission yield a size for the SO$_2$ of $\simeq
0\farcs8$--$1\farcs0$ (see Table~\ref{Tmdyn}). This size can also be
independently obtained by comparing the main beam  brightness temperature of
the low and high angular resolution maps of these two lines. For a Gaussian
distribution of the emission, the ratio of the low to high angular resolution
line intensity is $T_{\rm mb}^{\rm high}/T_{\rm mb}^{\rm low} =  (\theta_{\rm
low}^2 + \Omega^2)/(\theta_{\rm high}^2 + \Omega^2)$,  where $\Omega$ is the
source size, $\theta_{\rm high}$ and $\theta_{\rm low}$ are the beam sizes of
the high and low angular resolution maps, respectively.  Table~\ref{Tradex}
shows the intensities of the 13$_{2,12}$-13$_{1,13}$ and 
16$_{3,13}$-16$_{2,14}$ lines for the low and high angular resolution maps.
Using this method, the SO$_2$ source size is slightly smaller, 
$0\farcs6\pm0\farcs1$ (Table~\ref{Tradex}). Therefore, we adopt a  SO$_2$ disk
size in the range of $0\farcs5$--$1\farcs0$ (425--850~AU). This implies that
the brightness temperatures of the 13$_{2,12}$-13$_{1,13}$ and 
16$_{3,13}$-16$_{2,14}$ lines are $\simeq 62$--84~K and 68--94~K,
respectively.  The area within the solid lines of Figure~\ref{Radex} shows the
possible solutions that yield an SO$_2$ 13$_{2,12}$-13$_{1,13}$ brightness
temperature within this range. 
\end{itemize}

By combining the two aforementioned criteria and using the 99\% confidence
interval of the $\chi^2$, the possible reasonable solutions get significantly
reduced. Thus, for example, $N\mathrm{(SO_2)}=1 \times 10^{18}$\cmd\ has better
$\chi^2$ solutions than lower values of the SO$_2$ column density, but these
solutions predict a brightness temperature significantly higher than the
expected value. The best solutions are for a kinetic temperature of $T_{\rm
kin}=120$--160~K and a volume density of $n\mathrm{(H_2)}\ga 2\times10^9$\cmt.
These values are, approximately, in agreement  with those derived from the dust
emission associated with MM1 disk (Paper I). The possible SO$_2$ column
densities are in the  4--$6\times 10^{17}$\cmd\ range.

%% For a beam of 0.68x.38 arcsec (dust HR map:
%% Nbm(H2) = 1.13e23 cm-2
%%  SO2 source size of 0.5-1.0 arcsec: ff = 0.492 / 0.795
%% Nbm(SO2) = N(SO2)*ff = 1.5-3.2e17 cm-2 
%% X(SO2) = 1.3-2.8e-8
The abundance of the SO$_2$ emission can be estimated from the derived SO$_2$
column density taking into account the filling factor of the observations.
In Paper I, we used the 1.4~mm dust emission at an angular resolution of
$\simeq0\farcs51$ to derive a beam averaged gas column density of
$N\mathrm{(H_2)}\simeq1.1\times10^{25}$\cmd. For this angular resolution the
filling factor of a source with a $0\farcs5$--$1\farcs0$ size is 0.49--0.80.
Therefore, the SO$_2$ beam averaged column density at this angular resolution is
2.0--$4.8\times10^{17}$\cmd, which implies a SO$_2$ abundance of
$X\mathrm{[SO_2]}\simeq1.7$--$4.2\times10^{-8}$.

%% Tmb(SO 78-77)=0.5+-0.1K Beam=5.072arcsec ff=0.00962-0.0370 for
%% 0.5-1.0arcsec size
%% Tmb(SO 55-44)=2.3+-0.04K
%% Tb(SO 78-77)=13.5-52.0 K => N(SO)=8e16-3.5e17cm-2 => Nba(SO)=0.39-2.8e17cm-2
%% Tb(SO 55-44)=62-239 K =>    slightly contaminated ...
%% X(SO)=3.5e-9-2.5e-8

\section{DISCUSSION}
\subsection{A rotating molecular disk/ring toward MM1}
\subsubsection{Evidence for a disk}
The emission of the SO$_2$, SO and H$_2$CO at the position of MM1 seems to 
consist of a compact rotating circumstellar structure, probably a rotating
gaseous disk or ring (Figs. \ref{Fcubos} and \ref{Fmoms}). All
the species detected with the SMA toward MM1 are expected to be present in a
$\sim150$~K disk, in which a central protostar is evaporating the gas from the
grain mantles (e.g., \citealt{1997Charnley,2004Maret}). In addition,
the SO$_2$ abundance (2--4$\times10^{-8}$) obtained with
RADEX is consistent with that found in the innermost parts of the envelopes of
hot cores, where the temperature is greater than 100~K
(\citealt{2003VanderTak}).

The case for a rotating disk surrounding the protostar in MM1 becomes stronger 
in light of the new results reported here. The velocity gradient within the 
molecular emission of the SO$_2$ lines appears compact, with a radius
R$\simeq$425--850~AU, surrounding the compact (R$<150$~AU) 1.4~mm dusty  disk
and the thermal radio jet (Fig.~\ref{Fposter}). Furthermore, the position angle
between the extreme blue and redshifted channels of two SO$_2$ lines (see
\S3.4),  is roughly perpendicular to the radio jet axis, with a position angle
110--130$\degr$. 

\subsubsection{Dynamical mass}
We now estimate the dynamical mass with the 50\% contour of the \lsbii
and \usbii position-velocity diagrams in Fig. \ref{Fmoms}. We assume an
equilibrium between the centrifugal and gravitational forces and also that 
the disk is seen edge--on. 
We  use the virial equation, $M_{\rm dyn}=(v^2 R)/G$, which requires that
the dominant force is due to gravity. This gives a dynamical mass of
15$\pm5$\msun and 11$\pm3$\msun (Table \ref{Tmdyn}) by using the \lsbii and 
the \usbii data. The virial approximation is valid 
when the system is spherically symmetric regardless of its density 
distribution. It is in addition a good
approximation for thick disks (as appears to be the case of MM1, \S5.1.3).

The continuum observations toward MM1 (Paper I) have shown a very 
compact disk--like structure ($R<150$~AU, 4.1\msun) and a more extended and 
weaker envelope ($R\sim1500$~AU, 1.5\msun). On the other hand the SO$_2$ 
rotating structure has a radius of 425--850~AU. For this radius range the 
gravitational effect of the compact disk--like structure could be considered 
as part of the central mass, therefore, using the virial equation is a 
reasonable approximation to find out the protostar plus inner--disk mass.

To probe the impact of the extended molecular structure on the 
centrifugal balance of forces in MM1, we consider an additional term due to
a flattened disk, representing the SO$_2$ disk--like structure. We assume that
this structure has a mass of 1.5\msun and a radius of 1500~AU, the 
characteristics 
of the dusty envelope of MM1. We also assume a disk surface density inversely 
proportional to the distance from the protostar ($\rho(R)\propto R^{-1}$). 
In addition, we include another term to the balance of forces due to
a pressure gradient $(1/\rho)(dP/dR)$ (see \citealt{1986Torrelles}).
To estimate this pressure term we use either a sound speed of 0.7\kms 
(corresponding to a 150~K gas temperature), or a turbulent plus thermal 
random velocity of $\sim1$\kms. 
Both terms (extended disk and pressure gradient) could induce a departure 
from the Keplerian behavior of the disk. However, we find that both are one 
order of magnitude smaller than that of the central mass. Thereby,
the central mass obtained is essentially the same as that estimated with the 
virial approximation. 
% To make a better estimate of the balance of forces in MM1, two terms 
% could be added to the virial equation of the disk:  (i)
% a gravitational force produced by an extended and flattened envelope (see 
% Paper I) with a
% surface density inversely proportional to the distance from the protostar
% ($\rho(R)\propto R^{-1}$) and (ii) a force due to a pressure gradient
% $(1/\rho)(dP/dR)$ (see \citealt{1986Torrelles}). These terms could induce a
% departure from the Keplerian behavior of the disk. However, we find that 
% the two terms are one order of magnitude
% smaller than that of the central mass (to estimate the pressure term we use
% a sound speed of 0.7\kms, for a temperature of 150~K, or a turbulent plus 
% thermal random velocity of $\sim1$\kms). We have thus estimated the central
% mass (protostellar mass plus inner disk-like mass) using the \lsbii and the 
% \usbii data as 15$\pm5$\msun and 11$\pm3$\msun, respectively 
% (Table \ref{Tmdyn}). 
% from the pure virial estimation.
%\end{itemize}
% \\
% {\it LA SIGUIENTE FRASE QUEDA FUERA DE CONTEXTO CON LO QUE SE ACABA DE DECIR:
% ACTUALIZALA}
% \\
% Taking into account the data from the \lsbii and the \usbii data we finally 
% estimate a dynamical mass of about 11--15\msun

If the SO$_2$ rotating structure is not edge--on (i.e.,
i$\neq90\degr$), the derived total mass is just a lower limit. Nevertheless,
the edge--on approximation seems adequate for the MM1 disk, as the thermal
radio jet is one of the largest jets ($\sim5.3$~pc in projected distance)
observed up to now. Therefore, it is expected to be close to the plane of the
sky (\citealt{1993Marti}). However, \cite{1987Yamashita} estimated an inclination
between 60$\degr$ and 74$\degr$ by modeling the geometry of the infrared
reflection nebula, which would increase the dynamical mass to a range of
14--20\msun.

\subsubsection{Stability of the disk}
Due to the high mass of the compact dusty disk (about 4\msun), it is reasonable
to ask whether the molecular disk is stable against gravitational disturbances.
If we take as the disk parameters T$=120$--160~K, N$=1\times10^{25}$\cmd (and
hence the surface density, $\Sigma=N\cdot\mu\cdot\mathrm{m_H}=40$~g\cmd, using
a mean molecular weight of  2.3 for molecular hydrogen and helium at 10~K), and
the disk velocity 3.4--4.3\kms at 876--740~AU (Table \ref{Tmdyn}) from the
center of the disk, we can approximate the value of the Toomre parameter
(\citealt{1964Toomre}), defined as $$Q=\frac{\mathrm{c_s}\cdot\kappa}{\pi\cdot
G\cdot\Sigma}\quad.$$ In the equation above, $\mathrm{c_s}$ is the sound
velocity, $\kappa$ is the epicyclic frequency (which can be approximated as
$\Omega$ for Keplerian motion or as $2\Omega$ for solid body rotation, $\Omega$
being the angular velocity), G is the gravitational constant and $\Sigma$ is
the surface density. If Q is less than 1 then the disk is unstable. First, we
estimate the Toomre parameter using the thin disk assumption (e.g.,
\citealt{2006Kratter}), and a high epicyclic frequency (that of solid body
rotation). The resulting value is $Q=0.42$--0.72. It appears that the molecular
disk would be unstable even in the unlikely case of a rigid body rotation. 

However, \cite{2007Cesaroni} (see also \citealt{2001Durisen}) pointed out that
the thickness of disks should have a great stabilizing effect on these.
Following their approximation for thick disks by assuming a disk thickness
H$=\mathrm{c_s^2}/(\pi G\Sigma)=35$--47~AU, a disk mass of
$\mathrm{M_d}=4$\msun, and a total mass in a wide range of
$\mathrm{M_{tot}}=10$--20\msun, we obtain
Q$=\sqrt{2\cdot(\mathrm{H/R})\cdot(\mathrm{M_{tot}}/\mathrm{M_d}})=0.40$--0.80,
which is again smaller than unity. 

Although the Q$=0.8$ case is closer to unity, the MM1 disk--like structure
seems therefore probably unstable. In such a case, the gravitational
perturbations can lead to the appearance of dense spiral arms, rings, arcuate
structures or even dense fragments, which could result in the formation of a
clumpy structure (\citealt{2001Durisen,2007Durisen}).

\subsubsection{Mass of the protostar}
We can also roughly estimate the mass of the protostar at the position of MM1. 
If we assume an edge--on disk (which provides again with a lower limit), since 
the gas plus dust mass of the MM1 disk was estimated from the 1.4~mm millimeter 
continuum emission (see Paper I) to be about 4$\pm$1\msun, the mass of the 
protostar would be about 7$\pm$4--11$\pm$6\msun. Although the estimation has 
an uncertainty larger than a 50\%, this agrees well with the expected mass for a
$>3300$\lsun (Paper I) protostar, which could be related to a $\sim10$\msun
ZAMS star (Table 5 of \citealt{1998Molinari}).

\subsubsection{Absorption of disk emission by a foreground cloud}
The blueshifted emission of the  \lsbiv and \formal\ lines  from the disk
appears to be missing (Fig. \ref{Fcubos}). One possibility is that the interferometer
has filtered out some extended flux. However one would expect some remnant emission 
from the rotating structure. Another possibility is that gas from the
foreground component of the cloud between 8.5 and 11.8\kms (Figs.
\ref{Fsocube} and \ref{Fformalcube}) may have absorbed the blueshifted emission from
the disk.  Firstly, the extended dense molecular component traced by the C$^{17}$O
and H$_2$CO appears blueshifted at the same velocity range where the missing
velocity disk component is observed in  \lsbiv and \formal. And secondly, the
other SO transition, \lsbi, shows only compact emission toward MM1, as in the
other SO$_2$ lines. This SO line and also the SO$_2$ lines, have a higher
excitation, therefore they can only be excited in the hot environment around the disk.

\subsection{Emission from the cavity walls\label{cavity}}
Figs. \ref{Fmomso} and \ref{Fsocube} show the SO emission extending about
$10\arcsec$--$15\arcsec$ to the north of the position of MM1. The SO emission
appears clumpy, with some of the clumps probably associated with the molecular
outflows of the region.  In particular, the \lsbiv line emission,  and also in
a lesser measure the H$_2$CO, seems also to surround the thermal radio jet
axis, with a geometry reminiscent of a V-shape. A possibility is that part of
the emission of these lines is tracing the walls of the outflow cavity
excavated by the thermal radio jet originated at the position of MM1 (as that
of outflow cavities associated with low-mass protostars; see e.g.,
\citealt{2000Lee, 2006Arce, 2007Jorgensen, 2009SantiagoGarcia}).  This
quiescent molecular emission (i.e., showing narrow lines) associated with 
powerful outflows could be photoilluminated by the UV radiation generated in
the shocks  within the outflows, as has been observed in several HH objects 
(\citealt{2002Girart,2005Girart,2003Viti,2006Viti}). Indeed, in the case of
the  HH~80--81 and HH~80N system, there is a shock--induced  photodissociation
region along all the thermal radio jet, reported by \cite{2001Molinari}. 

\subsection{Is the Molecular Core actually a hot core?}

The only molecular lines clearly detected toward the molecular core (MC)
northwest of MM2 are \lsbiv and \formal. This source has been previously
identified as a hot molecular core (\citealt*{2009Qiu}). Both lines show a
narrow component (e.g., the FWHM of the SO line is $\sim 2.5$\kms) centered at 
\vlsr $\simeq$10--12\kms, and a redshifted wing spreading up to about 30\kms.
The narrower component is related to the quiescent gas in the molecular core,
while the component with the wider velocity range is probably due to shocked
gas entrained by the outflows of the region. Herein, we discuss the possible
origin of the MC. 

It has been proposed that the MC is a hot core surrounding a massive protostar.
This is based on the detection of hot core molecular tracers, such as CH$_3$OH,
HNCO, and OCS \citep{2009Qiu}. However, we note that the temperature estimated
in the MC is $\simeq 45$~K \citep{2009Qiu} which is about a factor of 2 lower 
than the temperature found in hot cores. In addition, the lack of dust emission
associated with the MC ($\la 10$~mJy at 1.4~mm) implies that for the
aforementioned temperature the total mass of the MC has an upper limit of $\sim
0.25$\msun \citep{2009Qiu}, too small for a hot core. Could it be possible that
MC is a hot corino such as IRAS~16293--2422 \citep{2009Rao}? At the distance of
\ggd, IRAS~16293--2422 source B would have a flux density of 10~mJy at 1.4~mm,
below the upper limit of our observations. But, the MC temperature is still
lower than that of this hot corino and the intensity of the  observed molecular
lines in source IRAS~16293--2422 B would be far from being detected with the
current sensitivity at the distance of \ggd.

A possible alternative is that MC is tracing the interaction of the dense
molecular  core with a strong shock produced in a molecular outflow in a way
that the external  heating and the radiation of the gas produced a strong
outgassing phase  (\citealt{2006Viti}), exciting many molecular transitions, as
it has been observed in  other sources, such as Orion
(\citealt{2002Liu,2010Zapata_hc}; see also \citealt*{1996Chernin}). 

There is a class I CH$_3$OH maser (\citealt{2004Kurtz,2009Qiu}) detected about
$1\arcsec$ south--east of MC. This class of masers is associated with shocked
molecular gas (\citealt{2004Kurtz}). The most likely scenario  is that the
molecular outflow from MM2 has encountered a molecular core at the MC
position.  The northwest lobe of this outflow is redshifted and its path is
crossed by the MC (see Figs.~\ref{Fsocube} and \ref{Fformalcube}). Therefore,
it seems possible that the outflow could partially hit the molecular core at
MC, producing a stationary bow shock which could entrain the molecular gas, as
found in some molecular outflows such as L1157 and IRAS~04166+2706
(\citealt{1995Zhang,2009SantiagoGarcia,2010Tafalla}). Given the location of the
MC, it is also possible that it is not an isolated core, but it forms part of
the cavity walls associated with the thermal radio jet (section \S 5.2).

\section{CONCLUSIONS}
We have performed an SMA observational study of the molecular gas toward 
the central region of \ggd, and have obtained the following main results.

\begin{itemize}
\item The \lsbii and \usbii data show a compact (R$\simeq425$--850~AU) molecular
disk--like structure spatially coincident with the position of the compact
millimeter dusty  source MM1 and the radio jet located at the \ggd position. 
The molecular structure is perpendicular to the radio jet axis and it shows a
clear velocity gradient that we interpret as rotation. The dynamical mass
(11--15\msun) allowed us to derive the mass of the protostar (7--11\msun),
making use of the  mass that was inferred from the continuum data (Paper I). 

The RADEX analysis constrains the disk--like structure temperature between 120
and 160~K  and its volume density as $\gtrsim2\times10^9$\cmt.  We have also
found that the rotating system could be unstable due to  gravitational
disturbances.

\item The \co emission shows an elongated structure of about
0.2~pc~$\times$~0.1~pc  with an orientation roughly perpendicular to the radio
jet. This emission  seems to belong to the dense core in which the protostars
associated with the MM1, MM2 and MC sources are forming.

\item The H$_2$CO and the SO emission appears clumpy and is mainly enhanced
toward  the position of MM1 and MC. It is possibly tracing in part the cavity
walls  excavated by the thermal radio jet. At the position of MC, it shows a
line profile with a red wing extending up to tens of \kms. This line profile is
probably due to shocked gas entrained by the outflows in the region. We
speculate that a possible scenario is the interaction between the outflow
associated with MM2 and the dense molecular core at this position.
\end{itemize}

\acknowledgments
We thank all members of the SMA staff that made these observations possible.
MFL acknowledges financial support from DGAP-UNAM, M\'exico. JMG are supported
by the Spanish MICINN AYA2008-06189-C03 and the Catalan AGAUR  2009SGR1172
grants. SC and MFL acknowledge support from CONACyT grant 60581, M\'exico. SC
and MFL thank the hospitality of the Institut de Ciencies de l'Espai
(CSIC-IEEC), Bellaterra, Catalunya, Spain. YG acknowledges support from
CONACyT grants 80769 and 49947-F.

\bibliography{biblio}

\newpage
%%%%%%%%%%%%%%%%%%%%%%%%%%%%%%%%%% TABLES %%%%%%%%%%%%%%%%%%%%%%%%%%%%%%%%%%%%%%

%%
%% RADEX Table 1

%%%%%%%%%%%%%%%%%%%%%%%% FIGURES %%%%%%%%%%%%%%%%%%%%%%%%%%%%%%%%%%%%%%%%%%%%%%%

\end{document}